\documentclass[twocolumn,showpacs,aps,prl,letterpaper]{revtex4}
\usepackage{graphicx}
\usepackage{dcolumn}
\usepackage{amsmath}
\usepackage{epsfig}
\long\def\inst#1{\par\nobreak\kern 4pt\nobreak
    {\itshape #1}\par\vskip 10pt plus 3pt minus 3pt}
\RequirePackage{xspace}
\usepackage{relsize}

\def\babar{\mbox{\slshape B\kern-0.1em{\smaller A}\kern-0.1em
    B\kern-0.1em{\smaller A\kern-0.2em R}}}
\def\Kbar    {\kern 0.18em\overline{\kern -0.18em K}{}\xspace}

\def\Kz      {\ensuremath{K^0}\xspace}
\def\Kzb     {\ensuremath{\Kbar^0}\xspace}
\def\KzKzb   {\ensuremath{\Kz {\kern -0.16em \Kzb}}\xspace}

\def\Ks     {\ensuremath{K_S}\xspace}
\def\Kl     {\ensuremath{K_L}\xspace}
\def\KsKs   {\ensuremath{\Ks {\kern -0.16em \Ks}}\xspace}
\def\KlKl   {\ensuremath{\Kl {\kern -0.16em \Kl}}\xspace}
\def\KsKl   {\ensuremath{\Ks {\kern -0.16em \Kl}}\xspace}
\def\KlKs   {\ensuremath{\Kl {\kern -0.16em \Ks}}\xspace}
\def\Dbar    {\kern 0.18em\overline{\kern -0.18em D}{}\xspace}
\def\Dz      {\ensuremath{D^0}\xspace}
\def\Dzb     {\ensuremath{\Dbar^0}\xspace}
\def\DzDzb   {\ensuremath{\Dz {\kern -0.16em \Dzb}}\xspace}
\def\Bbar    {\kern 0.18em\overline{\kern -0.18em B}{}\xspace}

\def\Bz      {\ensuremath{B^0}\xspace}
\def\Bzb     {\ensuremath{\Bbar^0}\xspace}
\def\BzBzb   {\ensuremath{\Bz {\kern -0.16em \Bzb}}\xspace}
\def\Bu      {\ensuremath{B^+}\xspace}
\def\Bub     {\ensuremath{B^-}\xspace}

\def\BpBm    {\ensuremath{\Bu {\kern -0.16em \Bub}}\xspace}

\newcommand{\optbar}[1]{\shortstack{{\tiny (\rule[.4ex]{1em}{.1mm})}
  \\ [-.7ex] $#1$}}
\def\BorBbar    {\kern 0.18em\optbar{\kern -0.18em B}{}\xspace}
\def\DorDbar    {\kern 0.18em\optbar{\kern -0.18em D}{}\xspace}
\def\KorKbar    {\kern 0.18em\optbar{\kern -0.18em K}{}\xspace}
\def\CP                {\ensuremath{C\!P}\xspace}
\def\pep2{PEP-II}
\mathchardef\Upsilon="7107
\def\Y#1S{\ensuremath{\Upsilon{(#1S)}}\xspace}

\begin{document}

\title{
\large \bfseries \boldmath $\Dz-\Dzb$ mixing  in $\Upsilon(1S)
\rightarrow \Dz \Dzb$ decay at Super-B}

\author{Hai-Bo Li}\email{lihb@mail.ihep.ac.cn}
\author{Mao-Zhi Yang}\email{yangmz@mail.ihep.ac.cn}
\affiliation{ CCAST (World Laboratory), P.O.Box 8730, Beijing 100080, China  \\
 Institute of High Energy Physics, P.O.Box 918, Beijing  100049, China }


\date{\today}


\begin{abstract}
$\Dz-\Dzb$ mixing and significant $CP$ violation in the charm
system may indicate the signature of new physics. In this study,
we suggest that the coherent $\DzDzb$ events from the decay of
$\Upsilon(1S) \rightarrow \Dz \Dzb$ can be used to measure both
mixing parameters and $CP$ violation in charm decays. The neutral
$D$ mesons from $\Upsilon(1S)$ decay are strongly boosted, so that
it will offer the possibility to measure the proper-time interval,
$\Delta t$,  between the fully-reconstructed $\Dz$ and $\Dzb$.
Both coherent and time-dependent information can be used to
extract $\Dz-\Dzb$ mixing parameters. The sensitivity of the
measurement should be improved at B factories or a super-B.

\end{abstract}


\maketitle


Due to the smallness of the Standard Model (SM) $\Delta C=0$
amplitude, $\Dz-\Dzb$ mixing offers a unique opportunity to probe
flavor-changing interactions which may be generated by new
physics. The most promising place to produce $\DzDzb$ pairs with
low backgrounds is the $\psi(3770)$ resonance just above the
$\DzDzb$ threshold. The current experiments, such as CLEO-c and
BESIII~\cite{besiii}, are all symmetric $D$ meson factories, on
which the time information cannot be used. It is very hard to
build an asymmetric $\tau$-Charm factory in order to separate the
two $\Dz$ decay vertices since we need a strong boost of the $D$
meson to measure the mixing parameters.  Although the
time-dependent analyses have been done at $B$ factories, the $D$
mesons produced there are incoherent. In $\Upsilon(1S) \rightarrow
\DzDzb$ decay, both $D$ mesons are strongly boosted in the rest
frame of the $\Upsilon(1S)$ with the Lorentz boost factor of
($(\beta \gamma)_D = 2.33$), precise determination of the
proper-time interval ($\Delta t$) between the two $D$ meson decays
is available. Both coherence and time information are essential to
measure the $\Dz- \Dzb$ mixing and $CP$ violation.

In this paper, we consider the possible observations of $\Dz
-\Dzb$ mixing and $CP$ violation in the $\Upsilon(1S) \rightarrow
\DzDzb$ decay, in which the coherent $\DzDzb$ events are generated
with strong boost. Here we assume that possible strong multiquark
effects that involve seaquarks play no role in $\Upsilon(1S) \to
\DzDzb$ decays~\cite{voloshin}. The $\Upsilon(1S)$ decays will
provide another opportunity to search for $\Dz- \Dzb$ mixing and
understand the source of \CP violation in charm system. The
amplitude for $\Upsilon(1S)$ decaying to $\DzDzb$ is $\langle
\DzDzb|H|\Upsilon(1S) \rangle$, and the $\DzDzb$ pair system is in
a state with charge parity $C= -1$, which can be defined
as~\cite{petrov_2005}
\begin{equation}
|\DzDzb\rangle^{C=-1} = \frac{1}{\sqrt{2}} \left [ |\Dz \rangle
|\Dzb\rangle - |\Dzb\rangle |\Dz\rangle\right ]. \label{eq:d0d0}
\end{equation}
Although there is a weak current contribution in $\Upsilon(1S)
\rightarrow \DzDzb$ decay, which may not conserve charge parity,
the $\DzDzb$ pair can not be in a state with $C= +1$. The reason
is that the relative orbital angular momentum of $\DzDzb$ pair
must be $l=1$ because of angular momentum conservation. A
boson-pair with $l=1$ must be in an anti-symmetric state, the
anti-symmetric state of particle-anti-particle pair must be in a
state with $C= -1$.

We shall analyze the time-evolution of $\DzDzb$ system produced in
$\Upsilon(1S)$ decay.

In the assumption of $CPT$ invariance, the weak eigenstates of
$\Dz - \Dzb$ system are $|D_L\rangle = p
|\Dz\rangle+q|\Dzb\rangle$ and $|D_H\rangle = p |\Dz\rangle
-q|\Dzb\rangle$ with eigenvalues $\mu_L = m_L -
\displaystyle\frac{i}{2} \Gamma_L$ and $\mu_H = m_H -
\displaystyle\frac{i}{2} \Gamma_H$, respectively, where the $m_L$
and $\Gamma_L$ ($m_H$ and $\Gamma_H$) are the mass and width of
``light (L)" $\Dz$ ( ``heavy (H)" $\Dz$) meson. Following the
$\Upsilon(1S) \rightarrow \DzDzb$ decay, the $\Dz$ and $\Dzb$ will
go separately and the proper-time evolution of the particle states
$|\Dz_{\small\mbox{phys}}(t)\rangle$ and
$|\Dzb_{\small\mbox{phys}}(t)\rangle$ are given by
\begin{eqnarray}
|\Dz_{\small\mbox{phys}}(t) \rangle & = & g_+(t)  |\Dz \rangle
-\frac{q}{p}
g_-(t) |\Dzb \rangle, \nonumber \\
 |\Dzb_{\small\mbox{phys}}(t) \rangle & = & g_+(t) |\Dzb \rangle - \frac{p}{q} g_-(t) |\Dz \rangle,
\label{eq:d0_time}
\end{eqnarray}
where
\begin{eqnarray}
g_{\pm} = \frac{1}{2} (e^{-im_H t-\frac{1}{2} \Gamma_H t} \pm
e^{-im_L t - \frac{1}{2}\Gamma_L t} ),
 \label{eq:define}
\end{eqnarray}
with definitions
\begin{eqnarray}
m &\equiv& \frac{m_L + m_H}{2}, \, \, \Delta m \equiv m_H - m_L,
\nonumber \\
\Gamma &\equiv & \frac{\Gamma_L + \Gamma_H}{2}, \, \Delta \Gamma
\equiv \Gamma_H - \Gamma_L,
 \label{eq:define2}
\end{eqnarray}
Note that here $\Delta m$ is positive by definition, while the
sign of $\Delta \Gamma$ is to be determined by experiments.

 In practice, one define the following mixing parameters
\begin{eqnarray}
 x \equiv \frac{\Delta m}{\Gamma}, \, y \equiv \frac{\Delta
 \Gamma}{2\Gamma}.
\label{eq:define3}
\end{eqnarray}
Then we consider a $\DzDzb$ pair in $\Upsilon(1S)$ decay with
definite charge-conjugation eigenvalue. The time-dependent wave
function of $\DzDzb$ system with $C=-1$ can be written as
\begin{eqnarray}
|\DzDzb (t_1,t_2) \rangle &=& \frac{1}{\sqrt{2}} [
|\Dz_{\small\mbox{phys}}({\bf k_1},t_1)
\rangle|\Dzb_{\small\mbox{phys}}({\bf k_2}, t_2) \rangle \nonumber \\
 & - & |\Dzb_{\small\mbox{phys}}({\bf k_1}, t_1) \rangle |\Dz_{\small\mbox{phys}}({\bf k_2}, t_2) \rangle ],
 \label{eq:D0D0_time}
\end{eqnarray}
where ${\bf k_1}$ and ${\bf k_2}$ are the three-momentum vector of
the two $D$ mesons.  We now consider decays of these correlated
system into various final states. An early study of correlated
$\DzDzb$ decays in to specific flavor final states, at a
$\tau$-Charm factory, was carried out by Bigi and
Sanda~\cite{bigi_1986}. Xing~\cite{xing_1997} had considered
time-dependent decays into correlated pairs of states at
$\psi(3770)$ and $\psi(4140)$ peaks. The amplitude of such joint
decays, one $D$ decaying to a final state $f_1$ at proper time
$t_1$, and the other $D$ to $f_2$ at proper time $t_2$, is given
by~\cite{xing_1997}
\begin{eqnarray}
A(\Upsilon(1S)&\to&
\Dz_{\small\mbox{phys}}\Dzb_{\small\mbox{phys}} \to f_1
f_2) \equiv
\frac{1}{\sqrt{2}}\times \nonumber \\
&& \{a_-[g_-(t_1)g_+(t_2)-g_+(t_1)g_-(t_2)] + \nonumber \\
&& a_+[g_-(t_1)g_-(t_2)-g_+(t_1)g_+(t_2)]\}
 \label{eq:amp}
\end{eqnarray}
where
\begin{eqnarray}
a_+ &\equiv& \overline{A}_{f_1}A_{f_2} - A_{f_1}\overline{A}_{f_2}
=A_{f_1}A_{f_2}\frac{p}{q}(\lambda_{f_1}-\lambda_{f_2}), \nonumber \\
a_- &\equiv& \frac{p}{q} A_{f_1}A_{f_2}
-\frac{q}{p}\overline{A}_{f_1}\overline{A}_{f_2} =
A_{f_1}A_{f_2}\frac{p}{q}( 1-\lambda_{f_1}\lambda_{f_2}),\nonumber \\
 \label{eq:aa}
\end{eqnarray}
with $A_{f_i} \equiv \langle f_i|{\cal H}|\Dz\rangle$,
$\overline{A}_{f_i} \equiv \langle f_i|{\cal H}|\Dzb\rangle$, and
define
\begin{eqnarray}
\lambda_{f_i} \equiv
 \frac{q}{p}\frac{\langle f_i | {\cal H}|\Dzb\rangle}{\langle f_i|{\cal
 H}|\Dz\rangle} =\frac{q}{p}\frac{\overline{A}_{f_i}}{A_{f_i}},
\label{eq:lambda_define}
\end{eqnarray}
\begin{eqnarray}
\overline{\lambda}_{\overline{f}_i} \equiv
 \frac{p}{q}\frac{\langle \overline{f_i} | {\cal H}|\Dz\rangle}{\langle \overline{f_i}|{\cal
 H}|\Dzb\rangle}
 =\frac{p}{q}\frac{A_{\overline{f_i}}}{\overline{A}_{\overline{f_i}}}.
\label{eq:lambda_define2}
\end{eqnarray}


 In the process $e^+e^- \to \Upsilon(1S) \to \DzDzb$ the central-of-mass energy is
far above the threshold of $\DzDzb$ pairs, so that the decay-time
difference (t=$\Delta t_- = (t_2 - t_1)$) between
$\Dz_{\small\mbox{phys}} \to f_1$ and $\Dzb_{\small\mbox{phys}}
\to f_2$ can be measured easily. From Eq.~(\ref{eq:amp}),  one can
derive the general expression for the time-dependent decay rate,
in agreement with~\cite{PDG2006}:
\begin{eqnarray}
&&\frac{d\Gamma(\Upsilon(1S)\to
\Dz_{\small\mbox{phys}}\Dzb_{\small\mbox{phys}} \to f_1
f_2)}{dt} =  {\cal N} e^{-\Gamma|t|}\times\nonumber \\
&&[(|a_+|^2+|a_-|^2)\mbox{cosh}(y\Gamma t) + (|a_+|^2 -
|a_-|^2)\mbox{cos}(x\Gamma t) \nonumber \\
&& -2{\cal R}e(a^*_+ a_-)\mbox{sinh}(y\Gamma t) +2{\cal I}m(a^*_+
a_-)\mbox{sin}(x\Gamma t)],
 \label{eq:decay_rate}
\end{eqnarray}
where ${\cal N}$ is a common normalization factor, in
Eq.~(\ref{eq:decay_rate}), terms proportional to $|a_+|^2$ are
associated with decays that occur without any net oscillation,
while terms proportional to $|a_-|^2$ are associated with decays
following a net oscillation. The other terms are associated with
the interference between these two cases. In the following
discussion, we define
\begin{eqnarray}
R(f_1,f_2; t) \equiv \frac{d\Gamma(\Upsilon(1S)\to
\Dz_{\small\mbox{phys}}\Dzb_{\small\mbox{phys}} \to f_1 f_2)}{dt}.
 \label{eq:decay_rate_define}
\end{eqnarray}

The time-dependent rate expression simplifies if one of the states
(say, $f_2$) is a $CP$ eigenstate $S_{\eta}$ with eigenvalue $\eta
= \pm$:
\begin{eqnarray}
  R(f_1,S_\eta; t) &=& {\cal N} |A_{S_{\eta}}|^2 |A_{f_1}|^2
e^{-\Gamma| t|}\times\nonumber \\
&&[2|\lambda_{f_1} +\eta|^2\mbox{cosh}(y\Gamma t) - \nonumber \\
&& 2 \eta(|\lambda_{f_1} + \eta |^2)\mbox{sinh}(y\Gamma t)],
 \label{eq:decay_rate_cp}
\end{eqnarray}
where $A_{S_\eta} = \langle S_\eta |{\cal H}| \Dzb\rangle$, and we
have used $CP|\Dz \rangle = - |\Dzb\rangle$ and $\lambda_{S_\eta}=
- \eta=\mp $ by neglecting $CP$ violation in decay, $\Dz-\Dzb$
mixing and the interference of the decay with and without mixing.

 Now we consider the following cases for the $D$ meson decays
 to various final states, such as semileptonic, hadronic, and $CP$
 eigenstates.

 (1)  $(l^-X^+,K^+\pi^-;  t)$:
\begin{eqnarray}
 R(l^- X^+,K^+\pi^-;  t) &=& {\cal N}|A_l|^2 |\bar{A}_{K^+\pi^-}|^2\left|\frac{q}{p}\right|^2 e^{-\Gamma |t|}\times \nonumber \\
 &&((1+|\overline{\lambda}_{K^+ \pi^-}|^2)\mbox{cosh}(y \Gamma t) \nonumber \\
&-& (1-|\overline{\lambda}_{K^+\pi^-}|^2)\mbox{cos}(x\Gamma t) \nonumber \\
&+& 2{\cal R}e(\overline{\lambda}_{K^+\pi^-})\mbox{sinh}(y\Gamma t) \nonumber \\
&+& 2{\cal I}m(\overline{\lambda}_{K^+\pi^-})\mbox{sin}(x\Gamma
t)).
 \label{eq:decay_lk_1}
\end{eqnarray}

(2)  $(l^+X^-,K^-\pi^+;  t)$:
\begin{eqnarray}
 R(l^+ X^-,K^-\pi^+;  t) &=& {\cal N}|A_l|^2 |A_{K^-\pi^+}|^2 \left|\frac{p}{q}\right|^2 e^{-\Gamma |t|}\times \nonumber \\
 &&((1+|\lambda_{K^- \pi^+}|^2)\mbox{cosh}(y \Gamma t) \nonumber \\
&-& (1-|\lambda_{K^-\pi^+}|^2)\mbox{cos}(x\Gamma t) \nonumber \\
&+& 2{\cal R}e(\lambda_{K^-\pi^+})\mbox{sinh}(y\Gamma t) \nonumber \\
&+& 2{\cal I}m(\lambda_{K^-\pi^+})\mbox{sin}(x\Gamma t)).
 \label{eq:decay_lk_2}
\end{eqnarray}

(3) $(l^+X^-,K^+\pi^-;  t)$:
\begin{eqnarray}
 R(l^+ X^-,K^+\pi^-;  t) &=& {\cal N} |A_l|^2 |\bar{A}_{K^+\pi^-}|^2 e^{-\Gamma |t|}\times \nonumber \\
 &&((1+|\overline{\lambda}_{K^+ \pi^-}|^2)\mbox{cosh}(y \Gamma t) \nonumber \\
&+& (1-|\overline{\lambda}_{K^+\pi^-}|^2)\mbox{cos}(x\Gamma t) \nonumber \\
&+& 2{\cal R}e(\overline{\lambda}_{K^+\pi^-})\mbox{sinh}(y\Gamma t) \nonumber \\
&-& 2{\cal I}m(\overline{\lambda}_{K^+\pi^-})\mbox{sin}(x\Gamma
t)).
 \label{eq:decay_lk_3}
\end{eqnarray}

(4)  $(l^-X^+,K^-\pi^+;  t)$:
\begin{eqnarray}
 R(l^- X^-,K^-\pi^+;  t) &=& {\cal N} |A_l|^2 |A_{K^-\pi^+}|^2 e^{-\Gamma |t|}\times \nonumber \\
 &&((1+|\lambda_{K^- \pi^+}|^2)\mbox{cosh}(y \Gamma t) \nonumber \\
&+& (1-|\lambda_{K^-\pi^+}|^2)\mbox{cos}(x\Gamma t) \nonumber \\
&+& 2{\cal R}e(\lambda_{K^-\pi^+})\mbox{sinh}(y\Gamma t) \nonumber \\
&-& 2{\cal I}m(\lambda_{K^-\pi^+})\mbox{sin}(x\Gamma t)).
 \label{eq:decay_lk_4}
\end{eqnarray}

(5)  $(l^\pm_1 X^\mp, l^\mp_2 X^\pm;  t)$:
\begin{eqnarray}
 R(l^\pm_1 X^\mp,l^\mp_2 X^\pm;  t) &=& {\cal N} |A_{l_1}|^2 |A_{l_2}|^2  e^{-\Gamma |t|}\times \nonumber \\
 &&(\mbox{cosh}(y \Gamma t)+ \mbox{cos}(x\Gamma t)),
 \label{eq:decay_ll}
\end{eqnarray}
where $l_1$ and $l_2$ could be electron or muon.

(6)  $(l^\pm X^\mp, S_\eta;  t)$:
\begin{eqnarray}
  R(l^- X^+,S_\eta; t) &=& R(l^+ X^-,S_\eta; t) \arrowvert_{q \Leftrightarrow p} \nonumber \\
 & =&{\cal N} |A_{l}|^2|A_{S_{\eta}}|^2
e^{-\Gamma| t|}\times\nonumber \\
&&[ 2 \mbox{cosh}(y\Gamma t) -2 \eta{\cal
R}e\left(\frac{q}{p}\right) \mbox{sinh}(y\Gamma
t) \nonumber \\
&& +2\eta{\cal I}m \left(\frac{q}{p}\right)\mbox{sin}(x\Gamma t)],
 \label{eq:decay_l_cp}
\end{eqnarray}
where $q \Leftrightarrow p$ indicates the exchange of $q$ and $p$, and $|q/p|=1$ is taken.

(7)  $(K^- \pi^+, S_\eta;  t)$:\\
 For this case, it is the same as the result in Eq.~(\ref{eq:decay_rate_cp})
 for $f_1 = K^-\pi^+$ when $CP$ violation is neglected.

(8) $(K^- \pi^+, K^+ \pi^-;  t)$: \\
 For the given final states $f_1
f_2= (K^-\pi^+)(K^+\pi^-)$, the situation becomes more
complicated, one can obtain the following expression after a
lengthy calculation:
\begin{eqnarray}
&& R(K^- \pi^+, K^+ \pi^-; t) = {\cal N}
|A_{K^-\pi^+}\overline{A}_{K^+\pi^-}|^2
e^{-\Gamma| t|} \times \nonumber \\
&&[(|1-\lambda_{K^-\pi^+}\overline{\lambda}_{K^+\pi^-}|^2+|\lambda_{K^-\pi^+}
- \overline{\lambda}_{K^+\pi^-}|^2)\mbox{cosh}(y\Gamma t) \nonumber \\
&&+
(|1-\lambda_{K^-\pi^+}\overline{\lambda}_{K^+\pi^-}|^2-|\lambda_{K^-\pi^+}
- \overline{\lambda}_{K^+\pi^-}|^2)\mbox{cos}(x\Gamma t) \nonumber \\
&&+2({\cal
R}e(\lambda_{K^-\pi^+}\overline{\lambda}_{K^+\pi^-}-1){\cal
R}e(\lambda_{K^-\pi^+} - \overline{\lambda}_{K^+\pi^-}) +\nonumber
\\
&& {\cal I}m(\lambda_{K^-\pi^+}\overline{\lambda}_{K^+\pi^-}){\cal
I}m(\lambda_{K^-\pi^+} - \overline{\lambda}_{K^+\pi^-})
)\mbox{sinh}(y\Gamma t) \nonumber \\
&&- 2({\cal
R}e(\lambda_{K^-\pi^+}\overline{\lambda}_{K^+\pi^-}-1){\cal
I}m(\lambda_{K^-\pi^+} - \overline{\lambda}_{K^+\pi^-}) -\nonumber
\\
&&{\cal I}m(\lambda_{K^-\pi^+}\overline{\lambda}_{K^+\pi^-}){\cal
R}e(\lambda_{K^-\pi^+} - \overline{\lambda}_{K^+\pi^-})
)\mbox{sin}(x
\Gamma t) ]. \nonumber \\
 \label{eq:decay_kpkp}
\end{eqnarray}

(9) $(K^- \pi^+, K^- \pi^+;  t)$:
\begin{eqnarray}
 &&R(K^- \pi^+, K^- \pi^+;  t) = {\cal N}
|A_{K^-\pi^+}|^4|\frac{p}{q}|^2 e^{-\Gamma|t|}\nonumber \\
&&\times|\lambda_{K^-\pi^+}^2-1|^2[\mbox{cosh}(y\Gamma t)-
\mbox{cos}(x\Gamma t)].
  \label{eq:decay_nine}
\end{eqnarray}
Mixing is the necessary condition for this process to occur.

(10) $(K^+ \pi^-, K^+ \pi^-;  t)$:
\begin{eqnarray}
 &&R(K^+ \pi^-, K^+ \pi^-;  t) = {\cal N}
|\overline{A}_{K^+\pi^-}|^4|\frac{q}{p}|^2 e^{-\Gamma|t|}\nonumber \\
&&\times|\overline{\lambda}_{K^+\pi^-}^2-1|^2[\mbox{cosh}(y\Gamma
t)- \mbox{cos}(x\Gamma t)]
  \label{eq:decay_ten}
\end{eqnarray}

(11) $( l^+_1 X^-, l^+_2 X^-, t)$:
\begin{eqnarray}
 R(l^+_1 X^-, l^+_2 X^-, t) &=& {\cal N}e^{-\Gamma|t|} \left|
 \frac{p}{q} \right |^2 |A_{l^+_1 X^-}|^2 |A_{l^+_2 X^-}|^2 \nonumber \\
 &\times& \left (\mbox{cosh}(y\Gamma t) -\mbox{cos}(x\Gamma t) \right).
  \label{eq:decay_eleven}
\end{eqnarray}

 (12) $( l^-_1 X^+, l^-_2 X^+, t)$:
\begin{eqnarray}
  R(l^-_1 X^+, l^-_2 X^+, t) &=& {\cal N}e^{-\Gamma|t|} \left|
 \frac{q}{p} \right |^2 |\overline{A}_{l^-_1 X^+}|^2 |\overline{A}_{l^-_2 X^+}|^2 \nonumber \\
 &\times& \left (\mbox{cosh}(y\Gamma t) -\mbox{cos}(x\Gamma t) \right).
  \label{eq:decay_tel}
\end{eqnarray}

 In deriving the above formulas from Eqs.~(\ref{eq:decay_lk_1})
to~(\ref{eq:decay_tel}), we have assumed that: (1) $\Delta Q =
\Delta C$ rule holds,  $A_{l^-} = \langle l^- X^+|{\cal H}|\Dz
\rangle = \overline{A}_{l^+} =\langle l^+ X^-|{\cal H}|\Dzb
\rangle =0$; (2) $CPT$ invariance holds. The results in
Eqs.(~\ref{eq:decay_lk_1}), (\ref{eq:decay_lk_2})
(\ref{eq:decay_lk_3}) and (\ref{eq:decay_lk_4}) are in agreement
with those in Ref.~\cite{xing_1997}.

In order to simplify the above formula, we make the following
definitions:
\begin{eqnarray}
\frac{q}{p} \equiv (1+A_M) e^{-i\beta},
 \label{eq:qp_par}
\end{eqnarray}
where $\beta$ is the weak phase in mixing and $A_M$ is a
real-valued parameter which indicates the magnitude of $CP$
violation in the mixing, and for $f = K^- \pi^+$, we define
\begin{eqnarray}
\frac{\overline{A}_{K^-\pi^+}}{A_{K^-\pi^+}} \equiv - \sqrt{r}
e^{-i \alpha}; \,\, \frac{A_{K^+\pi^-}}{\overline{A}_{K^+\pi^-}}
\equiv - \sqrt{r^\prime} e^{-i \alpha^\prime},
 \label{eq:double_par}
\end{eqnarray}
where $r$ and $\alpha$ ($r^\prime$ and $\alpha^\prime$) are the
ratio and relative phase of the doubly Cabibbo-suppressed (DCS)
decay rate and the Cabibbo-favored (CF) decay rate. Then,
$\lambda_{K^-\pi^+}$ and $\overline{\lambda}_{K^+\pi^-}$ can be
parameterized as
\begin{eqnarray}
\lambda_{K^-\pi^+} = -\sqrt{r}(1+A_M) e^{-i(\alpha+\beta)}\, ,
 \label{eq:lambda_para}
\end{eqnarray}
\begin{eqnarray}
\overline{\lambda}_{K^+\pi^-} = -\sqrt{r^\prime} \frac{1}{1+A_M}
e^{-i(\alpha^\prime -\beta)}
 \label{eq:lambda_para2}
\end{eqnarray}
In order to demonstrate the $CP$ violation in decay, we define
$\displaystyle \sqrt{r} \equiv \sqrt{R_D} \frac{1}{1+A_D}$ and
$\displaystyle \sqrt{r^\prime} \equiv \sqrt{R_D}(1+A_D)$. Thus,
Eqs. (\ref{eq:lambda_para}) and (\ref{eq:lambda_para2}) can be
expressed as
\begin{eqnarray}
\lambda_{K^-\pi^+} =
-\sqrt{R_D}\frac{1+A_M}{1+A_D} e^{-i(\delta+\phi)}\, ,
 \label{eq:lambda_para_s}
\end{eqnarray}
\begin{eqnarray}
\overline{\lambda}_{K^+\pi^-} = -\sqrt{R_D} \frac{1+A_D}{1+A_M}
e^{-i(\delta -\phi)}\, ,
 \label{eq:lambda_para2_s}
\end{eqnarray}
where $\displaystyle\delta = \frac{\alpha +\alpha^\prime}{2}$ is
the averaged phase difference between DCS and CF processes, and
$\displaystyle\phi = \frac{ \alpha-\alpha^\prime}{2}+\beta$.

We can characterize the $CP$ violation in the mixing amplitude,
the decay amplitude, and the interference between mixing and
decay, by real-valued parameters $A_M$, $A_D$, and $\phi$ as in
Ref~\cite{nir_1999}. In the limit of $CP$ conservation, $A_M$,
$A_D$ and $\phi$ are all zero.  $A_M = 0$ means  no $CP$ violation
in mixing, namely, $|q/p|=1$; $A_D=0$ means no $CP$ violation in
decay, for this case, $r = r^\prime = R_D=
|\overline{A}_{K^-\pi^+}/A_{K^-\pi^+}|^2=|A_{K^+\pi^-}/\overline{A}_{K^+\pi^-}|^2$;
$\phi =0 $ means no $CP$ violation in the interference between
decay and mixing.

 Taking into account that $\lambda_{K^-\pi^+}$,
 $\overline{\lambda}_{K^+\pi^-} \ll 1$ and $x$, $y \ll 1$,
 keeping terms up to order $x^2$, $y^2$ and $R_D$ in the
expressions, neglecting $CP$ violation in  mixing,  decay and the
interference between decay with and without mixing ($A_M=0$,
$A_D=0$, and $\phi=0$), expanding the time-dependent for $x t$, $y
t \lesssim \Gamma^{-1}$, we can write the results from Eqs.
(\ref{eq:decay_lk_1}) to (\ref{eq:decay_tel}) as

(1)  $(l^-X^+,K^+\pi^-;  t)$:
\begin{eqnarray}
&& R(l^- X^+,K^+\pi^-;  t) = {\cal N}|A_l|^2 |\bar{A}_{K^+\pi^-}|^2 e^{-\Gamma |t|}\times \nonumber \\
 &&(2R_D - 2\sqrt{R_D} y^\prime \Gamma t  + R_M\Gamma^2 t^2 ),
 \label{eq:decay_lk_1_s}
\end{eqnarray}
where $\displaystyle R_M \equiv \frac{x^2 + y^2}{2}$ is  mixing
rate, and $y^\prime \equiv y \mbox{cos} \delta - x \mbox{sin}
\delta$.

(2) $(l^+X^-,K^-\pi^+;  t)$:
\begin{eqnarray}
 && R(l^+ X^-,K^-\pi^+;  t) = {\cal N}|A_l|^2 |A_{K^-\pi^+}|^2  e^{-\Gamma |t|}\times \nonumber \\
 && (2R_D - 2\sqrt{R_D} y^\prime \Gamma t + R_M \Gamma^2 t^2),
 \label{eq:decay_lk_2_s}
\end{eqnarray}
since the mixing in neutral $D$ is tiny, it is much more likely
that $x^2$, $y^2 \ll R_D$, cases (1) and (2) can be used for
measuring $R_D$.

 (3)  $(l^+X^-,K^+\pi^-;  t)$:
\begin{eqnarray}
 && R(l^+ X^-,K^+\pi^-;  t) = {\cal N} |A_l|^2 |\bar{A}_{K^+\pi^-}|^2 e^{-\Gamma |t|}\times \nonumber \\
 &&(2 - 2\sqrt{R_D}(y \mbox{cos}(\delta) + x \mbox{sin}(\delta)) \Gamma
 t \nonumber \\
&& + \frac{y^2 - x^2}{2}\Gamma^2 t^2).
 \label{eq:decay_lk_3_s}
\end{eqnarray}

(4)  $(l^-X^+,K^-\pi^+;  t)$:
\begin{eqnarray}
 && R(l^- X^-,K^-\pi^+;  t) = {\cal N} |A_l|^2 |A_{K^-\pi^+}|^2 e^{-\Gamma |t|}\times \nonumber \\
 &&(2 -2\sqrt{R_D}(y \mbox{cos}(\delta) + x \mbox{sin}(\delta)) \Gamma
 t \nonumber \\
&& + \frac{y^2 - x^2}{2}\Gamma^2 t^2) = R(l^+ X^-,K^+\pi^-;  t).
 \label{eq:decay_lk_4_s}
\end{eqnarray}
In the limit of no $CP$ violation, case (3) is the same as (4).

(5)  $(l^\pm_1 X^\mp, l^\mp_2 X^\pm;  t)$:
\begin{eqnarray}
 && R(l^\pm_1 X^\mp,l^\mp_2 X^\pm;  t) = {\cal N} |A_{l_1}|^2 |A_{l_2}|^2  e^{-\Gamma |t|}\times \nonumber \\
 && (2 +\frac{y^2 - x^2}{2}\Gamma^2 t^2).
 \label{eq:decay_ll_s}
\end{eqnarray}

(6)  $(l^\pm , S_\eta;  t)$:
\begin{eqnarray}
 &&  R(l^\pm, S_\eta; t) = {\cal N} |A_{l}|^2|A_{S_{\eta}}|^2
e^{-\Gamma| t|}\times\nonumber \\
&& (2 - 2 \eta ( y \mbox{cos} \beta \mp  x \mbox{sin}\beta) \Gamma
t + y^2 \Gamma^2 t^2),
 \label{eq:decay_l_cp_s}
\end{eqnarray}
where $y$ may be determined because phase $\beta=
arg[(V_{us}V^*_{cs}/(V_{cs}V^*_{us})] \sim 0$.

(7)  $(K^- \pi^+, S_\eta;  t)$:
\begin{eqnarray}
 &&  R(K^- \pi^+ , S_\eta; t) = {\cal N} |A_{K^-\pi^+}|^2|A_{S_{\eta}}|^2
e^{-\Gamma| t|}\times\nonumber \\
&&(\eta-\sqrt{R_D}\mbox{cos}\delta)^2(1-\eta y\Gamma t+
\frac{1}{2}y^2 (\Gamma t)^2),
 \label{eq:kpi_cp_s}
\end{eqnarray}
where $\mbox{cos}\delta$ can be measured in this case by combing
$\eta = -1$ and $+1$ final states.

(8)  $(K^- \pi^+, K^+ \pi^-;  t)$:
\begin{eqnarray}
&& R(K^- \pi^+, K^+ \pi^-; t) = {\cal N}
|A_{K^-\pi^+}\overline{A}_{K^+\pi^-}|^2
e^{-\Gamma| t|} \times \nonumber \\
&& (2 + \frac{y^2 -x^2}{2} \Gamma^2 t^2 - 4 R_D
\mbox{cos}(2\delta)).
 \label{eq:decay_kpkp_s}
\end{eqnarray}

(9)  $(K^- \pi^+, K^- \pi^+;  t)$:
\begin{eqnarray}
&& R(K^- \pi^+, K^- \pi^+;  t) = {\cal N} |A_{K^-\pi^+}|^4  e^{-\Gamma |t|}\times \nonumber \\
 &&(1-2R_D\mbox{cos}(2\delta))\frac{x^2+y^2}{2}\Gamma^2 t^2.
 \label{eq:decay_xnine s}
\end{eqnarray}
This process is proportional to mixing rate $R_M$, and can be used
to measure the mixing parameter directly.

(10)  $(K^+ \pi^-, K^+ \pi^-;  t)$: \\
  The result is the same as
$R(K^- \pi^+, K^- \pi^+;  t)$ when $CP$ violation is neglected.

(11)  $( l^+_1 X^-, l^+_2 X^-, t)$:
\begin{eqnarray}
 R(l^+_1 X^-, l^+_2 X^-, t) &=& {\cal N}e^{-\Gamma|t|} |A_{l^+_1 X^-}|^2 |A_{l^+_2 X^-}|^2 \times \nonumber \\
    && \frac{x^2+y^2}{2} \Gamma^2 t^2.
  \label{eq:decay_eleven_s}
\end{eqnarray}
One can also definitely measure the mixing rate in the like-sign
processes as in case (9) and (10).

 (12)  $( l^-_1 X^+, l^-_2 X^+, t)$: \\
 The result is the same as
  $( l^+_1 X^-, l^+_2 X^-, t)$ when $CP$ violation in mixing and decay is neglected.

 Note that in all the above cases, when $CP$ violation in decay is neglected, there is
$|A_{K^-\pi^+}|=|\bar{A}_{K^+\pi^-}|$.

\begin{figure}
\epsfig{file=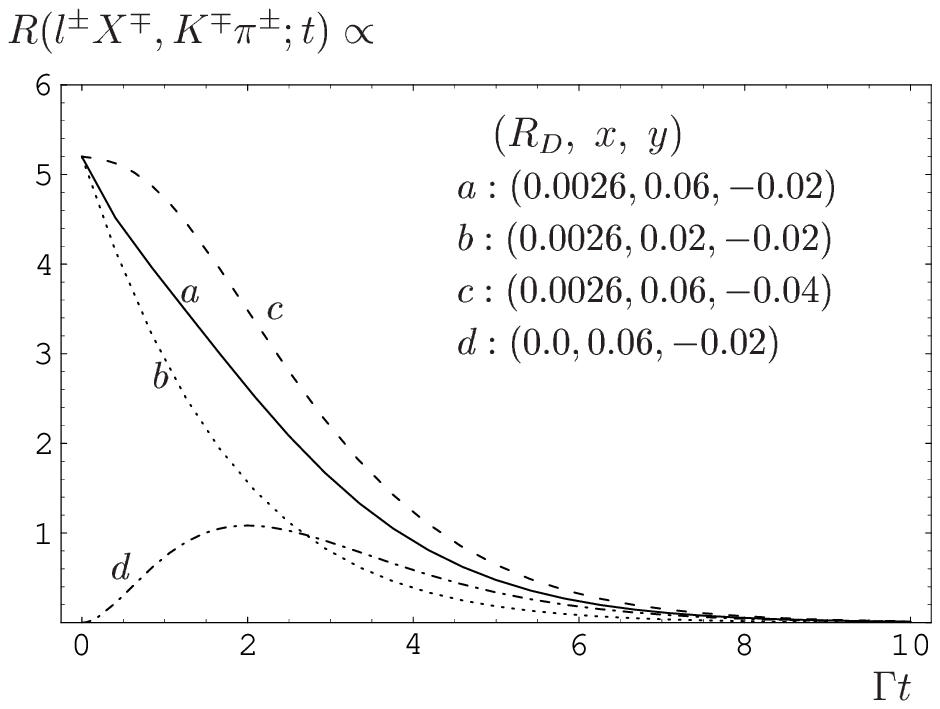,width=7cm,height=5cm} \caption{Illustrative
plot of the sensitivity of the decay rate $R(l^\pm
X^\mp,K^\mp\pi^\pm; t)$ to the mixing parameters of $x$, $y$ and
the ratio $R_D$, where $\delta=10^\circ$ is taken.}
\label{figure1}
\end{figure}

The experimental data from CLEO-c yield that the allowed values
for the mixing parameters $x$ and $y$ are: $y=-0.058\pm 0.066$,
$x<0.094$ \cite{PDG2006}. The ratio of the DCS decay rate to the
CF decay rate is $R_D\sim (V_{cs}V_{cd})^2\approx 0.0026$. The
illustrative plot of the decay rate $R(l^\pm X^\mp,K^\mp\pi^\pm;
t)$ is shown in Fig.\ref{figure1} by taking $\delta=10^\circ$. The
decay rate $R(l^\pm X^\mp,K^\mp\pi^\pm; t)$ is very sensitive to
the mixing parameters $x$, $y$ and the ratio of $R_D$ in the
region $\Gamma t\sim 1-6$. The other decay rates in Eqs.
(\ref{eq:decay_lk_3_s})-(\ref{eq:decay_kpkp_s}) are not sensitive
to the mixing parameters and $R_D$, because in these decay rates,
the $x$, $y$ and $R_D$ are only small correction to the dominant
contributions.

 In the Cabibbo-Kobayashi-Maskawa (CKM) framework $CP$
violation in the neutral $D$ system is very small and can be
safely neglected. However extension of the SM could induce new
physics of $CP$ violation~\cite{bigi_1986,grossman_2006}. The most
likely sizable effect is a possible new $CP$ violation phase,
$\phi = \mbox{arg} (q\overline{A}/p A)$, occurring in the
interference between mixing and decay amplitudes. Thus, in the
presence of $CP$ violation in the interference, we can construct
the following $CP$ observable based on the previous calculations.
We can look at the difference between the cases (3) and (4), and
define:
\begin{eqnarray}
A^{+-}_{CP} (t) \equiv \frac{R(l^+ X^-, K^+\pi^-) - R(l^- X^+,
K^-\pi^+)}{R(l^+ X^-, K^+\pi^-) + R(l^- X^+, K^-\pi^+)}.
 \label{eq:cp_inter}
\end{eqnarray}
As discussed in Ref~\cite{xing_1997}, the signal is due to the
interplay of DCS decay and mixing.  With help of
Eqs.~(\ref{eq:decay_lk_3}) and (\ref{eq:decay_lk_4}), we obtain
\begin{eqnarray}
A^{+-}_{CP} (t) = - \sqrt{R_D} ( y \mbox{sin}\delta - x
\mbox{cos}\delta ) \mbox{sin} \phi \times \Gamma t.
 \label{eq:cp_inter_values}
\end{eqnarray}
Here we keep both $x$ and $y$ terms since the current experiments
indicate that they may be at the same order, this is different
from Ref~\cite{xing_1997}.  The above asymmetry depends on
non-vanishing phase $\phi$, and also the mixing parameters. Within
the SM, it is
 of order ${\cal O}(10^{-3})$~\cite{bigi_charm2006}, which makes such
 an asymmetry unmeasurably tiny unless there is new
 physics~\cite{soni_2005}. By looking at the difference between
 $R(l^+, S_\eta; t)$ and $R(l^+, S_\eta; t)$ in case (6), we get
 the following asymmetry
\begin{eqnarray}
A^{S_{\eta}}_{CP} (t) \equiv \frac{R(l^+ , S_{\eta}) - R(l^-,
S_{\eta})}{R(l^+ , S_{\eta}) + R(l^-, S_{\eta})}.
 \label{eq:cp_eigenstates}
\end{eqnarray}
With the help of Eq. (\ref{eq:decay_l_cp}), we obtain:
\begin{eqnarray}
A^{S_{\eta}}_{CP} (t) = \eta x \mbox{sin}\beta \times \Gamma t,
 \label{eq:cp_eigenstates_value}
\end{eqnarray}
where $\beta$ is defined in Eq. (\ref{eq:qp_par}).  This $CP$ term
depends on the mixing parameter $x$ and the phase in the mixing
amplitude.

 In experiments, the KEK-B can move to $\Upsilon(1S)$ peak
without losing luminosity, more than 330$fb^{-1}$ data per year
could be taken~\cite{fuko}. The measured cross section at $\Upsilon(1S)$ peak
is $\sigma_R = 21.5 \pm 0.2 \pm 0.8$ nb by CLEO~\cite{CLEO_1983}.
One can estimate the total $\Upsilon(1S)$ events with one year
running of KEK-B are about $7.1\times 10^9$. While, at super-KEKB,
about $10^{11}$ $\Upsilon(1S)$ events can be obtained with one
year data-taking if the design luminosity is $8\times
10^{35}$~\cite{hazumi}. More recently, a super-B factory based on
the concepts of the Linear Collider (LC) is proposed~\cite{il_B},
the low energy beam and high energy beam are 4.0 and 7.0 GeV,
respectively.  The machine can run at both $\Upsilon(4S)$ and
$\Upsilon(1S)$ peaks with luminosity about $1.0 \times 10^{36}$.
Then about $10^{12}$ $\Upsilon(1S)$ events can be collected with
one year's run. In Ref.~\cite{qcd_1982}, one had estimated the
ratio of ${\cal BR}(\Upsilon(1S) \rightarrow D^+D^-)$ and ${\cal
BR}(\Upsilon(1S) \rightarrow K^+K^-)$ as
\begin{eqnarray}
\frac{{\cal BR}(\Upsilon(1S) \rightarrow D^+D^-)}{{\cal
BR}(\Upsilon(1S) \rightarrow K^+K^-)} \cong
\left(\frac{1}{0.2}\right)^2 \gg 1.0
 \label{eq:ratio}
\end{eqnarray}
where the current upper limit of ${\cal BR}(\Upsilon(1S)
\rightarrow K^+K^-)$ is $5.0\times 10^{-4}$ at 90\% confidence
level. One can expect that the decay rate of $\Upsilon(1S)
\rightarrow K^+K^-)$ is at the order of $10^{-6}$~\cite{qcd_1982},
so that we can estimate ${\cal BR}(\Upsilon(1S) \rightarrow
D^+D^-) \sim 10^{-4}
 - 10^{-5}$.  At the super-B factory, around $10^{7} - 10^{8}$
 $\DzDzb$ pairs can be collected in one year's data taking, which is
 comparable with that at the BESIII with four years integrated
 luminosity~\cite{li_2005}

It is known that one has to fit the proper-time distribution in
experiments to extract both the mixing and the $CP$ parameters, we
discuss the following two cases: (1) Case-I: at a symmetric
$\Upsilon(1S)$ factory, namely, the $\Upsilon(1S)$ is at rest in
the Central-Mass (CM) frame. Then, the proper-time interval
between the two $D$ mesons is calculated as
\begin{eqnarray}
  \Delta t = (r_{\Dz} -r_{\Dzb})\frac{m_D}{c{\bf |P|}},
 \label{eq:propertime}
\end{eqnarray}
where $r_{\Dz}$ and $r_{\Dzb}$ are $\Dz$ and $\Dzb$ decay length,
respectively, and ${\bf P}$ is the three-momentum vector of $\Dz$.
Since the momentum can be calculated with $\Upsilon(1S)$ decay in
the CM frame, all the joint $\DzDzb$ decays in this paper can be
used to study $D^0-\overline{D}^0$ mixing and the $CP$ violation
in the symmetric $\Upsilon(1S)$ factory. (2) Case-II: while, at an
asymmetric $\Upsilon(1S)$ factory,  the $\Upsilon(1S)$ will be
produced with a boost. In this case, the momentum of $\Dz$ and
$\Dzb$ will be different each other, and one has to fully
reconstruct at least one of the two $D$ mesons, since the the
proper-time interval between the two $D$ mesons is calculated as
\begin{eqnarray}
  \Delta t = r_{\Dz}\frac{m_D}{c{\bf |P_{\Dz}|}} -r_{\Dzb}\frac{m_D}{c{\bf |P_{\Dzb}|}},
 \label{eq:propertime_2}
\end{eqnarray}
where ${\bf |P_{\Dz}|}$ and ${\bf |P_{\Dzb}|}$ are momentum of
$\Dz$ and $\Dzb$, respectively.  In this case, the joint decays to
di-lepton in Eq.~(\ref{eq:decay_ll}) will not work, since both the
$D$ mesons cannot be fully reconstructed. One cannot obtain the
proper-time from such kind of experiment.

The average decay length of the $D^0$ meson in the rest frame of
$\Upsilon(1S)$ is $c\tau_{D^0} \times (\beta \gamma)_{D^0} \approx
290 \mu$m.  At B factories, such as Belle detector,  the impact
parameter resolution of vertex detector, which are directly
related to decay vertex resolution of $D^0$, are described in
Ref.~\cite{belle_svt}, from which we can get that the resolution
for the reconstructed $D^0$ decay length should be less than $100
\mu$m within the coverage of the detector. This means Belle
detector is good enough to separate
 the two $D^0$ decay vertices, so that the mixing parameters can
 be measured by using time information.

 All the results in this paper can also be applied to the
following processes,
\begin{eqnarray}
e^+e^- \rightarrow \Upsilon(1S) \rightarrow \pi^0 (\Dz \Dzb)_{C
=-1},
 \label{eq:another_pro}
\end{eqnarray}
where the $\Dz \Dzb$ pair are in a $C=-1$ state, for example,
$\Upsilon(1S) \rightarrow D^{*0} \Dzb \rightarrow \pi^0 \Dz \Dzb$.

 In conclusion, we suggest that the coherent $\DzDzb$ events
from the decay of $\Upsilon(1S) \rightarrow \Dz \Dzb$ can be used
to measure both the mixing parameters and the $CP$ violation in
charm decays. The neutral $D$ mesons from $\Upsilon(1S)$ decay are
strongly boosted, so that it will offer the possibility to measure
the proper-time interval, $\Delta t$,  between the
fully-reconstructed $\Dz$ and $\Dzb$. Both coherent and
time-dependent information can be used to extract $\Dz-\Dzb$
mixing parameters, in which  the sensitivity of the measurements
could be improved by comparing to the future measurement at the
BESIII with the same amount of $\DzDzb$ pairs.

One of the authors (H.~B.~Li) would like to thank Dr. Rong-Gang
Ping for helpful discussion. This work is supported in part by the
National Natural Science Foundation of China under contracts Nos.
10205017, 10575108, and the Knowledge Innovation Project of CAS
under contract Nos. U-612 and U-530 (IHEP).




\end{document}